\documentstyle[12pt]{article}
\begin{document}
\centerline{\Large\bf Maintaining a Wormhole with}
\vskip 0.08in
\centerline{\Large\bf a Scalar Field}
\vskip 1.0in
\centerline{Dan N. Vollick}
\centerline{Department of Physics}
\centerline{Okanagan University College}
\centerline{3333 College Way}
\centerline{Kelowna, B.C.}
\centerline{V1V 1V7}
\vskip 0.6in
\section*{\center{\bf Abstract}}
It is well known that it takes matter that violates the averaged weak 
energy condition to hold the throat of a wormhole open. The production
of such ``exotic'' matter is usually discussed within the context of
quantum field theory. In this paper I show that 
it is possible to
produce the exotic matter required to hold a wormhole open classically.
This is accomplished by coupling a scalar
field to matter that satisfies the weak energy condition. The energy-momentum
tensor of the scalar field and the matter separately satisfy the weak energy
condition, but there exists an interaction energy-momentum tensor that does
not. It is this interaction energy-momentum tensor that allows the wormhole to
be maintained.

\section*{\centerline{Introduction}}
A wormhole is a handle which connects two different space-times or two
distant regions in the same space-time \cite{Mo1,Mo2,Fr1}. 
To keep a wormhole open it is
necessary to thread its throat with matter that violates the averaged
weak energy condition \cite{Mo1,Mo2}. 
In other words, there exist null geodesics
passing through the wormhole, with tangent vectors $k^{\mu}=dx^{\mu}/
d\sigma$, which satisfy $\int_0^{\infty}T_{\mu\nu}k^{\mu}k^{\nu}d\sigma
<0$. Such matter obviously violates the weak energy condition which states
that $T^{\mu\nu}U_{\mu}U_{\nu}\geq 0$ for all non-spacelike vectors $U^{\mu}$.
The weak energy condition ensures that all observers will see a positive
energy density. Matter that violates the weak energy condition is called
exotic. Thus it takes exotic matter to hold a wormhole open.
   
Most discussions of exotic matter involve quantum field theory effects, 
such as the Casimir effect \cite{It1}. In this paper I show that it is
possible to generate the exotic matter required to maintain a wormhole
classically.
This is accomplished by
coupling a scalar field to matter which satisfies the weak energy condition.
The energy-momentum tensor of the scalar field and the matter separately 
satisfy the weak energy condition, but there exists an interaction 
energy-momentum tensor that does not. It is this interaction energy-momentum
tensor that allows the wormhole to remain open.
  
To create a wormhole I take two static, spherically symmetric, scalar-vac
solutions of the Einstein field equations and join them together. A
surface energy-momentum tensor will exist on the surface where these two
manifolds are joined. This surface energy-momentum tensor violates the
weak energy condition. However, if its source is a scalar field coupled to 
matter, I show that the energy-momentum tensor of the matter and of the field 
can both satisfy the weak energy condition. The violation of the 
weak energy condition for the total energy-momentum tensor is produced by
the interaction energy-momentum tensor. In addition to satisfying the
weak energy condition I show that the matter and scalar field also 
satisfy the dominant energy condition. The dominant energy condition
ensures that the four velocity associated with the local flow
of energy and momentum is non-spacelike. Thus a wormhole can be maintained 
classically
by coupling a scalar field to matter that satisfies the weak and dominant
energy conditions.
\section*{Equations of Motion and the
Energy-Momentum Tensor}
Consider a collection of charged time-like particles interacting with a scalar
field $\phi$. The action will be taken to be \cite{Vo1}.
\begin{equation}
S=-\Sigma_nm_n\int\sqrt{-g_{\mu\nu}U^{\mu}_nU^{\nu}_n}d\tau_n+\frac
{1}{2}\Sigma_n\int\lambda_n(\tau_n)[g_{\mu\nu}U^{\mu}_nU^{\nu}_n+1]
d\tau_n-\Sigma_n\alpha_n\int\phi(x_n(\tau_n))d\tau_n
\label{1}
\end{equation}
\begin{equation}
-\frac{1}{2}
\int\nabla^{\mu}\phi\nabla_{\mu}\phi\sqrt{g}d^4x
\end{equation}
where $x^{\mu}_n(\tau_n)$ and $U^{\mu}_n$ are the position and 
four-velocity of the $n^{th}$ particle, $\tau_n$ is the proper time along 
its world line, $m_n$ is its rest mass, 
$\lambda_n(\tau_n)$ are Lagrange 
multipliers, and the $\alpha_n$ are coupling constants.
  
The scalar field equations are found by varying the action with respect
to $\phi(x)$ and are given by
\begin{equation}
\Box^2\phi=\frac{1}{\sqrt{g}}\Sigma_n\alpha_n\int\delta^4(x^{\mu}-
x^{\mu}_n(\tau_n))d\tau_n,
\label{3a}
\end{equation}
where $\Box^2=\nabla^{\mu}\nabla_{\mu}$. 
The equations of motion for
the particles are found by varying the action with respect to $x^{\mu}_n
(\tau_n)$ and are given by
\begin{equation}
(m_n+\lambda_n)(\frac{dU^{\mu}_n}{d\tau_n}+\Gamma^{\mu}_{\alpha\beta}
U^{\alpha}_nU^{\beta}_n)+\frac{d\lambda_n}{d\tau_n}U^{\mu}_n=-\alpha_n
\nabla^{\mu}\phi.
\end{equation}
Contracting with $U^{\mu}_n$ gives
\begin{equation}
\frac{d\lambda_n}{d\tau_n}=\alpha_n\frac{d\phi}{d\tau_n}.
\end{equation}
Thus
\begin{equation}
\lambda_n=\alpha_n\phi.
\end{equation}
The equations of motion for the particles are then given by
\begin{equation}
\frac{d}{d\tau_n}[(m_n+\alpha_n\phi)U^{\mu}_n]+(m_n+\alpha_n\phi)\Gamma^{\mu}
_{\alpha\beta}U^{\alpha}_nU^{\beta}_n=-\alpha_n\nabla^{\mu}\phi.
\end{equation}
   
The energy-momentum tensor of the field and particles is given by
\begin{equation}
T^{\mu\nu}=\frac{2}{\sqrt{g}}\frac{\delta S}{\delta g_{\mu\nu}}.
\end{equation}
From (\ref{1}) the energy-momentum tensor is given by
\begin{equation}
T^{\mu\nu}=\Sigma_n\frac{1}{\sqrt{g}}\int(m_n+\alpha_n\phi)U^{\mu}_nU^{\nu}
_n\delta^4(x-x_n(\tau_n))d\tau_n+\nabla^{\mu}\phi\nabla^{\nu}\phi-
\frac{1}{2}g^{\mu\nu}\nabla_{\alpha}\phi\nabla^{\alpha}\phi
\label{emten}
\end{equation}
There is therefore an interaction energy-momentum tensor given by
\begin{equation}
T^{\mu\nu}_{(I)}=\Sigma_n\frac{\alpha_n}{\sqrt{g}}\int\phi(x_n)U^{\mu}_n
U^{\nu}_n\delta^4(x-x_n(\tau_n))d\tau_n.
\label{8}
\end{equation}
This interaction energy-momentum tensor is necessary if
\begin{equation}
\nabla_{\mu}T^{\mu\nu}=0
\end{equation}
is to give the correct equations of motion for the particles.
   
Now consider a collection of particles which all have the same
value of $\alpha/m$ (the simplest possibility would be to take $\alpha_n=
\pm m_n$).
From (\ref{3a}) and (\ref{emten}) it can be seen that it is the trace of the 
particle energy-momentum tensor which acts as the source of the scalar field.
In the continuum limit equations (\ref{3a}) and (\ref{emten}) become
\begin{equation}
\Box^2\phi=-\alpha^* T_m
\label{phi}
\end{equation}
and 
\begin{equation}
T^{\mu\nu}=(1+\alpha^*\phi)[(\rho+P)U^{\mu}U^{\nu}+Pg^{\mu\nu}]+
\nabla^{\mu}\phi\nabla^{\nu}\phi-\frac{1}{2}g^{\mu\nu}\nabla^{\alpha}\phi
\nabla_{\alpha}\phi,
\end{equation}
where
\begin{equation}
T_m=3P-\rho_m
\end{equation}
is the trace of the matter energy-momentum tensor, $\rho$ is the rest mass
density, $\alpha^*=\alpha/m$, and $P$ is the pressure.
\section*{\centerline{Wormhole Solutions}}
In this section exact wormhole solutions with matter energy-momentum 
tensors that satisfy the weak and dominant energy conditions will be found. 
The first step is to obtain static spherically symmetric
solutions to the scalar-vac Einstein field equations.
Two such solutions will then be joined at the radial coordinate $r=b$. In
the process of joining these manifolds together, surface energy and
stresses will be produced at $r=b$. The properties of the surface 
energy-momentum tensor will then be examined in relation to the various energy
conditions.
   
The metric will be taken to be of the form
\begin{equation}
ds^2=-e^{\alpha}dt^2+e^{\beta}[dr^2+r^2(d\theta^2+\sin^2\theta d\phi^2)].
\end{equation}
The Einstein field equations
\begin{equation}
R_{\mu\nu}=-8\pi G(T_{\mu\nu}-\frac{1}{2}g_{\mu\nu}T)
\end{equation}
give the three independent equations
\begin{equation}
\frac{1}{2}\alpha^{''}+\frac{1}{4}\alpha^{'}(\alpha^{'}+\beta^{'})+
\frac{\alpha^{'}}{r}=0
\label{1a}
\end{equation}
\begin{equation}
\frac{1}{2}\alpha^{''}+\beta^{''}-\frac{1}{4}\alpha^{'}(\beta^{'}-
\alpha^{'})+\frac{\beta^{'}}{r}=-8\pi G\phi^{'2}
\label{2a}  
\end{equation}
and
\begin{equation}
\frac{1}{2}\beta^{''}+\frac{1}{4}\beta^{'}(\alpha^{'}+\beta^{'})+
\frac{1}{2r}(3\beta^{'}+\alpha^{'})=0.
\label{333a}
\end{equation}
The scalar field equation is
\begin{equation}
\frac{\partial}{\partial r}\left(r^2e^{(\alpha+\beta)/2}\frac{\partial\phi}
{\partial r}\right)=0.
\label{scalar}
\end{equation}
Equation (\ref{1a}) plus equation (\ref{333a}) gives
\begin{equation}
\frac{1}{2}(\alpha+\beta)^{''}+\frac{1}{4}(\alpha+\beta)^{'2}+\frac{3}{2r}
(\alpha+\beta)^{'}=0.
\end{equation}
The general solution to this differential equation, which satisfies the
boundary condition $\alpha+\beta\rightarrow 0$ as $r\rightarrow\infty$ is
\begin{equation}
\alpha+\beta=2\ln\left(1-\frac{A}{r^2}\right),
\label{5a}
\end{equation}
where $A$ is an integration constant. Using this to eliminate $\beta^{'}$ in
(\ref{1a}) gives
\begin{equation}
\frac{1}{2}\alpha^{''}+\frac{A\alpha^{'}}{r(r^2-A)}+\frac{\alpha^{'}}{r}=0.
\label{6a}
\end{equation}
The solution of this equation is
\begin{equation}
\alpha=AD\int\frac{dr}{r^2-A}
\end{equation}
where D is an integration constant. The solution for $\alpha$ depends on
whether $A>0$, $A<0$, or $A=0$. It turns out that $A=0$ gives a flat space-time
and that there is no consistent solution for $A<0$. Thus I will only consider
the case $A>0$. The solution to (\ref{6a}) which satisfies the boundary
condition $\alpha\rightarrow 0$ as $r\rightarrow\infty$ is
\begin{equation}
\alpha=a\ln\left(\frac{1-\frac{Gm}{ar}}{1+\frac{Gm}{ar}}\right)
\end{equation}
where $a=\frac{1}{2}D\sqrt{A}$ and $Gm/a=\sqrt{A}$. 
The solution for $\beta$ can then
be found from (\ref{5a}) and the line element is given by
\begin{equation}
ds^2=-\left(\frac{1-Gm/ar}{1+Gm/ar}\right)^adt^2+\left(1+\frac{Gm}{ar}
\right)^4\left(\frac{1-Gm/ar}
{1+Gm/ar}\right)^{(2-a)}[dr^2+r^2(d\theta^2+\sin^2\theta d\phi^2)].
\label{7a}
\end{equation}
In the weak field limit, $|Gm/ar|<<1$,
\begin{equation}
g_{00}\simeq-\left(1-\frac{2Gm}{r}\right).
\end{equation}
Thus $m$ is the gravitational mass.
   
The scalar field can be found from (\ref{scalar}). The solution which satisfies the
boundary condition $\phi\rightarrow 0$ as $r\rightarrow\infty$ and is 
consistent with (\ref{2a}) is given by
\begin{equation}
\phi=\pm\sqrt\frac{4-a^2}{16\pi G}\ln\left(\frac{1-Gm/ar}{1+Gm/ar}\right)
\label{8a}
\end{equation}
with $-2\leq a\leq 2$.
Since (\ref{7a}) and (\ref{8a}) are invariant under $a\rightarrow -a$ only
$0\leq a\leq 2$ needs to be considered. Note that for $a=2$
the above solution reduces
to the Schwarzschild solution in isotropic coordinates.
   
The space-time geometry appears to be badly behaved at $r=Gm/a$
for $a\neq 2$. The Ricci scalar $R=g^{\mu\nu}R_
{\mu\nu}$ is given by
\begin{equation}
R=\frac{2G^2m^2r^4}{a^2}(4-a^2)(r+\frac{Gm}{a})^{-(4+a)}
(r-\frac{Gm}{a})^{(a-4)}
\end{equation}
which diverge as $r\rightarrow Gm/a$. The curvature scalar $I=R_{\mu\nu
\alpha\beta}R^{\mu\nu\alpha\beta}$ also diverges as $r\rightarrow
Gm/a$.
Thus $r=Gm/a$ is a physical
singularity in the space-time. In fact, it is a naked singularity since it
can be seen by distant observers. Thus, when joining the manifolds at $r=b$
it is necessary to take $b>Gm/a$.
  
Before joining the two scalar-vac manifolds together it is convenient
to change coordinates. Let $l$ be a new radial coordinate with $l=0$
at $r=b$ and
\begin{equation}
dl^2=\left(1+\frac{Gm}{ar}\right)^4\left(\frac{1-Gm/ar}{1+Gm/ar}
\right)^{(2-a)}dr^2.
\end{equation}
On one manifold take $0\leq l<\infty$ and on the other take 
$-\infty <l\leq 0$.
The manifold that consists of these manifolds joined at $l=0$ has
the line element
\begin{equation}
ds^2=-\left(\frac{1-Gm/ar}{1+Gm/ar}\right)^adt^2+dl^2+r^2
\left(1+\frac{Gm}{ar}\right)^4
\left(\frac{1-Gm/ar}{1+Gm/ar}\right)^{(2-a)}(d\theta^2+\sin^2\theta d\phi^2)
\label{line}
\end{equation}
where $r=r(l)$ and $-\infty <l<\infty$. The coordinates $r$ and $l$ are
related via
\begin{equation}
\frac{dr}{dl}=\pm\left(1+\frac{Gm}{ar}\right)^{-2}
\left(\frac{1-Gm/ar}{1+Gm/ar}\right)^{(a/2-1)}
\end{equation}
where the plus sign corresponds to the manifold with $l>0$ and the minus sign 
corresponds to the manifold with $l<0$. 
   
To find the surface energy-momentum tensor I will use the method developed
by Israel \cite{Is1,Is2}. The surface energy momentum tensor
\begin{equation}
S^{\mu}_{\;\;\nu}=\lim_{\epsilon\rightarrow 0}\int^{\epsilon}_{-\epsilon}
T^{\mu}_{\;\;\nu}dl
\label{9a}
\end{equation}
is given by
\begin{equation}
8\pi GS^{\mu}_{\;\;\nu}=\gamma^{\mu}_{\;\;\nu}-\delta^{\mu}_{\;\;\nu}
\gamma\;\;\;\;\;\;\;\;\;\;\;(\mu,\nu=0,2,3)
\end{equation}
where 
\begin{equation}
\gamma^{\mu}_{\;\;\;\nu}=K^{+\mu}_{\;\;\;\;\;\nu}-K^{-\mu}_{\;\;\;\;\;\nu}
\label{99a}
\end{equation}
$K^{+\mu}_{\;\;\;\;\;\nu}$ is the extrinsic curvature of the surface $r=b$
on the manifold with $l\geq 0$ and $K^{-\mu}_{\;\;\;\;\;\nu}$ is the extrinsic
curvature of the surface $r=b$ on the manifold with $l\leq 0$. For $\mu$ or
$\nu=1$, $S^{\mu}_{\;\;\nu}=0$. Using $K_{\mu\nu}=-\frac{1}{2}g_{\mu\nu,l}$
gives
\begin{equation}
S^t_{\;\; t}=\left(\frac{1}{2\pi Gb}\right)\left(1-\frac{G^2m^2}{a^2b^2}
\right)^{-1}\left(1+\frac{
Gm}{ab}\right)^{-2}\left(\frac{1-Gm/ab}{1+Gm/ab}\right)^{(a/2-1)}
\left(1-\frac{Gm}{b}+\frac{
G^2m^2}{a^2b^2}\right)
\label{S1}
\end{equation}
and
\begin{equation}
S^{\theta}_{\;\;\theta}=S^{\phi}_{\;\;\phi}=\left(\frac{1}{4\pi Gb}
\right)\left(1-\frac{
G^2m^2}{a^2b^2}\right)^{-1}\left(1+\frac{Gm}{ab}\right)^{-2}\left(
\frac{1-Gm/ab}{1+Gm/ab}\right)^{
(a/2-1)}\left(1+\frac{G^2m^2}{a^2b^2}\right).
\label{S2}
\end{equation}
In the previous section it was shown that the energy-momentum tensor for an
ideal fluid coupled to a scalar field is given by
\begin{equation}
T^{\mu\nu}=(1+\alpha^*\phi)T^{\mu\nu}_m+\nabla^{\mu}\phi\nabla^{\nu}\phi
-\frac{1}{2}g^{\mu\nu}\nabla_{\alpha}\phi\nabla^{\alpha}\phi
\label{alpha1}
\end{equation}
where $T^{\mu\nu}_m$ is the usual energy-momentum tensor for an ideal fluid.
The scalar field equation was also shown to be
\begin{equation}
\Box^2\phi=-\alpha^*T_{m}.
\label{beta1}
\end{equation}
The surface energy momentum tensor does not have the form of an ideal fluid
energy-momentum tensor because it contains only surface stresses (i.e.
$T^{\theta\theta},T^{\phi\phi}\neq 0$ but $T^{ll}=0$). For matter which
produces surface stresses I will  take (\ref{alpha1}) and
(\ref{beta1}) with
\begin{equation}
T^{\mu}_{m\;\;\nu}=[\sigma U^{\mu}U_{\nu}+S(\delta^{\mu}_{\;\;\theta}
\delta^{\theta}_{\;\;\nu}+\delta^{\mu}_{\;\;\phi}\delta^{\phi}_{\;\;\nu})]
\delta(l)
\end{equation}
where $\sigma$ is the surface density and $S$ is the surface stress.
Using (\ref{line}) and (\ref{9a})-(\ref{99a}) to find $S^{\mu}_{\;\;\nu}$
and equating it to (\ref{S1}) and (\ref{S2}) gives
\begin{equation}
(1+\alpha^*\phi)\sigma=-\left(\frac{1}{2\pi Gb}\right)\left(1-
\frac{G^2m^2}{a^2b^2}\right)^{-1}\left(1+
\frac{Gm}{ab}\right)^{-2}\left(\frac{1-Gm/ab}{1+Gm/ab}\right)^{(a/2-1)}
\left(1-\frac{Gm}{b}
+\frac{G^2m^2}{a^2b^2}\right)
\label{11a}
\end{equation}
and
\begin{equation}
(1+\alpha^*\phi)S=\left(\frac{1}{4\pi Gb}\right)\left(1-\frac{G^2m^2}{a^2b^2}
\right)^{-1}\left(1+
\frac{Gm}{ab}\right)^{-2}\left(\frac{1-Gm/ab}{1+Gm/ab}\right)^{(a/2-1)}
\left(1+\frac{G^2m^2}
{a^2b^2}\right).
\label{12a}
\end{equation}
For $0\leq a\leq 2$ it is easy to show that $(1+\alpha^*\phi)\sigma<0$
and $(1+\alpha^*\phi)S>0$.
   
Now consider the various energy conditions. 
An energy-momentum tensor
$T^{\mu\nu}$ will satisfy the weak energy condition if $T^{\mu\nu}U_{\mu}
U_{\nu}\geq 0$ for all non-spacelike vectors $U^{\mu}$. If the weak energy
condition is satisfied all observers will measure a positive energy density.
An energy momentum-tensor will satisfy the dominant energy condition if
it satisfies the weak energy condition and if $T^{\mu\nu}U_{\mu}$ is
non-spacelike for all non-spacelike vectors $U^{\mu}$. If the dominant
energy condition is satisfied all observers will measure the four vector
associated with the local flow of energy and momentum to be non-spacelike.
It is easy to show that the scalar field energy-momentum tensor satisfies
the weak and dominant energy conditions. 
    
The matter energy-momentum tensor $T^{\mu\nu}_m$ will satisfy the
weak energy condition iff \cite{Ha1,Wa1}
\begin{equation}
\sigma\geq 0\;\;\;\;\;\;\;\;\;\;\;\; and\;\;\;\;\;\;\;\;\;\;
\sigma+S\geq 0.
\end{equation}
Both of these conditions will be satisfied iff $1+\alpha^*\phi\leq 0$.
From (\ref{8a}) this gives
\begin{equation}
1\pm\alpha^*\sqrt{\frac{(4-a^2)}{16\pi G}}\ln\left(\frac{1+Gm/ab}{1-Gm/ab}
\right)
\leq 0.
\end{equation}
For a solution to exist
the lower sign must be chosen if $\alpha^*m>0$ and the upper sign must 
be chosen if $\alpha^*m<0$.
  
$T^{\mu\nu}_m$ will satisfy the dominant energy condition  iff it satisfies the
weak energy condition and if \cite{Ha1,Wa1}
\begin{equation}
\sigma\geq S
\end{equation}
This will be satisfied iff $1+\alpha^*\phi\leq 0$ and if
\begin{equation}
(1-\frac{Gm}{ab})^2+\frac{2Gm}{ab}(1-a)\geq 0.
\label{gamma1}
\end{equation} 
For $0\leq a\leq 1$ this will be satisfied for all $Gm/b$. For $1<a\leq 2$ this
will be satisfied for
\begin{equation}
-a<\frac{Gm}{b}\leq a^2-a\sqrt{a^2-1}.
\label{46}
\end{equation}
   
The final energy condition, which appears in some of the singularity theorems,
is the strong energy condition. For $\sigma>0$ the matter energy momentum tensor
will satisfy the dominant energy condition iff \cite{Ha1,Wa1}
\begin{equation}
\sigma +2S\geq 0
\end{equation}
From (\ref{11a}) and (\ref{12a}) it can be shown that the strong energy condition
will be satisfied iff $m\leq 0$. Since the strong energy condition does not
have a strong physical motivation, unlike the weak and dominant energy
conditions, I will not impose it on $T^{\mu\nu}_m$ (in fact the strong energy
condition can be violated by a massive classical scalar field{\cite{Ha1}).
   
Now consider the scalar field equation
\begin{equation}
\Box^2\phi=\alpha^*(\sigma-2S)\delta(l).
\end{equation}
Integrating from $l=-\epsilon$  to $l=\epsilon$ gives
\begin{equation}
\alpha^*(\sigma-2S)=\mp\sqrt{\frac{4-a^2}{\pi G}}\frac{Gm}{ab^2}
\left(1-\frac{G^2
m^2}{a^2b^2}\right)^{-1}\left(1+\frac{Gm}{ab}\right)^{-2}
\left(\frac{1-Gm/ab}{1+Gm/ab}\right)^{(a/2-1)}.
\label{50a}
\end{equation}
As before, if the weak energy condition is to be satisfied by $T^{\mu\nu}_m$,
the lower sign must be chosen if $\alpha^*m<0$ and the upper sign must be
chosen if $\alpha^*m<0$. Equation (\ref{50a}) is a constraint on $\sigma$
and $S$, which are given by (\ref{11a}) and (\ref{12a}).
Combining this with (\ref{11a}) and (\ref{12a}) gives
\begin{equation}
(1+\alpha^*\phi)=\pm\frac{\alpha^*(1-Gm/2b+G^2m^2/a^2b^2
)(Gm/ab)^{-1}}{\sqrt{\pi G(4-a^2)}}.
\end{equation}
Since $1-Gm/2b+G^2m^2/a^2b^2>0$ (for $0\leq a\leq 2$) any solution of this
equation will automatically satisfy the weak energy (choosing the upper or
lower sign as discussed above). This equation can be written as
\begin{equation}
f(x)=2x^2-ax+2\mp\Lambda\lambda x-\frac{1}{2}\Lambda^2x\ln
\left(\frac{1+x}{1-x}\right)=0
\label{51}
\end{equation}
where $\Lambda=\sqrt{4-a^2}$, $\lambda=\sqrt{4\pi G}/\alpha^*$, and $x=Gm/ab$.
For $0\leq a<2$, $\lim_{x\rightarrow\pm 1}f(x)=-\infty$ and $f(0)=2$. Thus
for $0\leq a<2$ there is at least one positive and one negative value of $x$,
in the interval $(-1,1)$, which satisfies $f(x)=0$. For $a=2$, $f(x)=2(
x^2- x+1)$ which has no real zeros. Thus, for $a=2$ it is not 
possible to join the 
two manifolds, independent of whether the energy conditions hold or not. 
This might seem strange
at first since the manifolds are Schwarzschild space-times, which
can be joined with the appropriate surface energy and stress. However, in
the usual Schwarzschild case there is no scalar field. In the case examined
here the vanishing of the scalar field gives the additional constraint $
\sigma-2S=0$, which is inconsistent with the energy and stress produced by
joining the Schwarzschild space-times. Thus for $0\leq a<2$ there exists 
at least one value of $Gm/ab$ that corresponds to a space-time in which
$T^{\mu\nu}_m$ satisfies the weak energy condition. If the range of $a$ is
restricted to $0\leq a\leq 1$ the corresponding $T^{\mu\nu}_m$ will also
satisfy the dominant energy condition. For $1<a<2$ equation (\ref{46}) must
be checked to see if $T^{\mu\nu}_m$ satisfies the dominant energy condition.
For $a=2$ it is not possible to join the manifolds together.
  
Before leaving this section I want to show that it is not possible for
$T^{\mu\nu}_m$ to satisfy the weak energy condition if the gravitational 
field described by (\ref{line}) is weak. In the weak field limit $x<<1$
and (\ref{51}) reduces to
\begin{equation}
(a\pm\lambda\Lambda)x\simeq 2.
\end{equation}
For $x$ to be small it is necessary that $|\lambda\Lambda|>>1$. Thus
$\pm\lambda\Lambda x>0$, which implies that
\begin{equation}
\pm\alpha^*m >0.
\end{equation}  
If $T^{\mu\nu}_m$ satisfies the weak energy condition the lower sign must
be chosen if $\alpha^*m>0$ and the upper sign must be chosen if $\alpha^*
m<0$. This is clearly inconsistent. Thus (\ref{line}) cannot describe a weak
gravitational field if $T^{\mu\nu}_m$ satisfies the weak energy condition.
In fact, it can be shown numerically that the smallest 
value of $|x|$ that satisfies (\ref{51}) ( with the appropriate sign) 
is $|x|\simeq 0.75$.
\section*{\centerline{Conclusion}}
The energy-momentum tensor for a scalar field coupled to an ideal fluid
was derived. In addition to the energy-momentum tensor for the matter and
the scalar field there exists an interaction energy-momentum tensor. The
interaction energy-momentum tensor can violate the weak and dominant energy
conditions even if the matter and scalar field energy-momentum tensors do not.
It is the interaction energy-momentum tensor that allows the wormhole to be
maintained.
   
A wormhole was created by joining two static, spherically symmetric,
scalar-vac solutions of the Einstein field equations. A surface
energy-momentum tensor that violates the weak energy condition exists on
the surface where the two space-times are joined. If the source of the
energy-momentum tensor is taken to be a scalar field coupled to matter
I showed that the energy-momentum tensor of the matter and scalar field
can satisfy the weak and dominant energy conditions. The violation of the
weak energy condition is produced by the interaction energy-momentum 
tensor.  Thus a wormhole can be maintained classically by coupling a 
scalar field to matter that satisfies the weak and dominant energy conditions.
Finally, I showed that it is not possible for the matter
energy-momentum tensor to satisfy the weak energy condition if the
gravitational field is weak.  

\end{document}